\documentclass[12pt]{revtex4}
\usepackage[dvips]{graphicx}
\usepackage{delarray}
\usepackage{amsmath}
\usepackage{amssymb}
\usepackage{here}
\usepackage{color}
\usepackage{url}
\newcommand{\ord}{\mathcal{O}}
\newcommand{\be}{\begin{equation}} \newcommand{\ee}{\end{equation}}

\newcommand{\vpa}{v_{\|}}
\newcommand{\vpe}{v_{\perp}}

\newcommand{\xpa}{x_{\|}}

\newcommand{\p}{\partial}

\newcommand{\red}[1]{\textcolor{black}{#1}}
\newcommand{\newred}[1]{\textcolor{black}{#1}}
\newcommand{\gyro}{{\sc gyro}}
\newcommand{\neo}{{\sc neo}}
\newcommand{\Ev}{\mathbf{E}}
\newcommand{\Bv}{\mathbf{B}}
\newcommand{\mN}{\mathcal{N}}
\newcommand{\energy}{\mathcal{E}}
\newcommand{\TEMII}{{\sc Case~1}}
\newcommand{\TEMIII}{{\sc Case~2}}
 
\begin{document}

\begin{center}
\Large

{\bf Impurity transport in trapped electron mode driven turbulence}\\
~\\*[0.5cm] \normalsize {A. Moll\'en$^1$,
  I. Pusztai$^{1,2}$,  T. F\"ul\"op$^1$, S. Moradi$^1$ \\
  \it\small $^1$ Department of Applied Physics, Nuclear Engineering,
  Chalmers University of Technology and Euratom-VR Association,
  G\"oteborg,
  Sweden}\\
{\it\small $^2$ Plasma Science and Fusion Center, Massachusetts
  Institute of Technology, Cambridge MA, USA.  }  \\ \today
\end{center}
\begin{abstract}
  Trapped electron mode turbulence is studied by
  gyrokinetic simulations with the \gyro~code \red{and an analytical
    model including the effect of a poloidally varying electrostatic
    potential}.  Its impact on radial transport of high-$Z$ trace
  impurities close to the core is thoroughly investigated and the
  dependence of the zero-flux impurity density gradient (peaking
  factor) on local plasma parameters is presented.  Parameters such as
  ion-to-electron temperature ratio, electron temperature gradient and
  main species density gradient mainly affect the impurity peaking
  through their impact on mode characteristics.  The poloidal
  asymmetry, the safety factor and magnetic shear have the strongest
  effect on impurity peaking, and it is shown that under certain
  scenarios where trapped electron modes are dominant, core
  accumulation of high-$Z$ impurities can be avoided. \red{We
    demonstrate that accounting for the momentum conservation property of the
    impurity-impurity collision operator can be important for an
    accurate evaluation of the impurity peaking factor.}

\end{abstract}

\maketitle

\section{Introduction}
Turbulence driven by unstable drift waves is considered to be
responsible for most of the observed cross-field particle and heat
transport in the core of tokamaks. In particular, ion gyro-radius
scale drift waves destabilized by the non-adiabatic response of
trapped electrons, the so-called trapped electron (TE) modes, 
\newred{
can play an important role, specifically in conditions where the 
electron heating power is large compared to the ion heating power 
and the electron temperature is larger than the ion temperature.
}
\newred{
They can also be important 
}
in transport barrier regions, where the density gradient is
large \cite{ernst}.

Since its original discovery
\cite{kadomtsev}, TE modes have been the topic of theoretical
investigations. They are usually categorized into the dissipative and
collisionless classes \red{\cite{fusionphysics12}}. 
The dissipative TE mode
requires a strong temperature gradient and large collisionality, while
the collisionless TE mode -- which is more likely to be destabilized
in reactor relevant conditions -- is driven by the electron curvature
drift resonance and can be destabilized even in the absence of
collisionality.  The collisionless TE mode can be driven purely by the
main species density gradient, 
\red{or} by the electron temperature gradient. 
Consequently it is customary to further divide the
collisionless TE mode into density gradient driven and electron
temperature gradient driven categories.
The stability and the turbulent fluxes driven by TE modes have been
analyzed in \cite{estrada,dannertjenko}. It has been shown in
\cite{dannertjenko,casati} that a quasilinear electrostatic
approximation might retain much of the relevant physics of TE mode
driven transport as it appears in nonlinear gyrokinetic simulations.
The purpose of this paper is to study the impurity transport driven by
TE modes.

It is well known that accumulation of impurities -- particularly those
with high charge number -- in the core of fusion plasmas has
debilitating effect on fusion reactivity due to radiative losses and
plasma dilution.  Results of fluid and gyrokinetic simulations
\cite{angioni2006,angioni2007,puiatti,fulopweiland,fulopnordman,braun,pusztai,moradi,bourdelle,parisot,villages,futatani,dubuit,Skyman2012,Howard2012,Nordman2011,angioni2009}
indicated that the anomalous impurity transport driven by
electrostatic microinstabilities in general, and TE modes in
particular, is determined by the competition of three main mechanisms:
curvature, thermodiffusion and parallel compressibility. The first of
these contribute to an inward impurity transport (when the magnetic
shear is positive), while thermodiffusion depends on the direction of
the mode propagation, being inward for modes propagating in the
electron diamagnetic direction such as the TE modes, however this
contribution is negligible for high-$Z$ impurities.  The sign of the
parallel compressibility contribution also depends on the direction of
the mode propagation, but is instead outward for TE modes and has a
charge to mass ratio dependence.

In recent years attention has been directed towards the role of TE
modes in impurity transport in plasmas with radio frequency (RF)
heating.  Various experiments reported reduced impurity accumulation
in such circumstances \cite{nue,dux,valisa}. In particular, it was
shown in Ref.~\cite{puiatti2003} that impurity transport was more
affected by the change in the plasma parameters due to RF heating than
by the generated sawtooth activity.  In
Ref.~\cite{angioni2006,angioni2007} it was argued that in a
TE-dominated ASDEX-U discharge with Electron Cyclotron Resonance
Heating (ECRH) the outward flows due to parallel compressibility
explained the reduction in the impurity density peaking. In Ion
Cyclotron Resonance Heating (ICRH) discharges on the JET tokamak the
differences in the steady state impurity density profiles under
minority heating (peaked impurity profiles) and mode conversion
heating (hollow or flat impurity profiles) were partially explained by
ITG and density gradient driven TE dominated transport, respectively
\cite{puiatti,giroudIAEA}. However, to be fully consistent with the
observations an assumption of a sub-dominant electron temperature
gradient driven TE mode was necessary. Further experimental studies in
JET plasmas \cite{valisa} showed a favorable impact of ICRH in
preventing the accumulation of metallic impurities accumulating in the
core. However, in this case a theoretical explanation based on the
presence of a TE mode driving outward impurity flux would be
unsatisfactory since these plasmas were ITG dominated.

Recently, a new possibility has emerged from the work reported in
\cite{fulop,moradi2,albert} where the observed outward directed
impurity flux is explained as an effect of poloidal asymmetries
generated by the ICRH.  The temperature anisotropy due to ICRH will
trap the minority heated ions on the low field side, leading to the
establishment of a poloidally varying ``equilibrium''
(i.e.~non-fluctuating) electrostatic potential (such asymmetries have
experimentally been demonstrated in \cite{reinke}). The
associated $\Ev\times \Bv$ drift acting as another degree of freedom
for impurities to respond to electrostatic perturbations modifies the
fluctuating impurity distribution.  It has been shown that under
experimentally relevant conditions the contribution of these
$\Ev\times \Bv$ drifts to the impurity particle transport can be
outward and might dominate the resulting steady state impurity
gradient even in ITG dominant regimes. 

The density peaking of high-$Z$ impurities in 
density or
temperature gradient driven TE dominated plasmas under RF induced
poloidal asymmetries is yet to be analyzed; this is the aim of the
present paper. Apart from numerical simulations with \gyro\
\cite{gyro} (mainly linear simulations, but a few nonlinear
simulations are also performed for comparison) an analytical model
including the effect of poloidal asymmetries \cite{albert,varenna} is
utilized.  The model is based on a solution to the linearized
gyrokinetic equation, and it suggests that the impurity velocity pinch
is governed by three separate contributions: one related to the
magnetic drifts (combined effects of curvature and thermodiffusion
pinch), another to the parallel impurity velocity pinch, and a third
part arising due the $\Ev\times \Bv$ drift in a poloidally varying
equilibrium electrostatic potential. \red{Using this model, we
    present a systematic comparison of impurity transport driven by
    density and temperature driven trapped electron modes,
    highlighting the effect of the parallel impurity motion,
    collisions, magnetic geometry (shear and safety factor) and
    poloidal asymmetries.}
 
The remainder of the paper is organized as follows.  In
Sec.~\ref{sec:stability}, we describe the baseline density and
temperature gradient driven TE mode cases and the linear stability
characteristics of them.  In Sec.~\ref{sec:impurity}, the density
peaking of high-$Z$ trace impurities is analyzed, and the dependence
on relevant plasma parameters, such as electron density and temperature gradients,
ion-to-electron temperature ratio, safety factor and magnetic shear,
is presented.  Impurity peaking factors are calculated in cases where
the impurities are poloidally symmetrically distributed, but also in
cases where a poloidally varying potential is present.  The results
are discussed and summarized in Sec.~\ref{sec:conclusions}.

\section{Stability}\label{sec:stability}
The TE mode instability is driven by the electron logarithmic temperature
gradient, $a/L_{Te}$, and/or the logarithmic density gradients,
$a/L_n$, whereas ITG modes are driven by the ion logarithmic
temperature gradient $a/L_{Ti}$. 
Here
$L_{n\alpha}=-\left[\partial\left(\ln{n_\alpha}\right)/\partial
  r\right]^{-1}$ and
$L_{T\alpha}=-\left[\partial\left(\ln{T_\alpha}\right)/\partial
  r\right]^{-1}$ represent the density and temperature scale lengths
of particle species $\alpha$, respectively, and $a$ the outermost
minor radius of the plasma.

In this paper we will study two baseline collisionless TE mode cases: one
driven by the density gradients and one driven by the electron
temperature gradient.  For the second case the ion temperature
gradient is set to zero, in order to obtain pure TE turbulence. This
represents a situation with dominant central electron heating.  Our
baseline cases have the following local profile and magnetic geometry
parameters:
\begin{description}
\item[] \TEMII: Density gradient driven TE mode \newline
  $R_0/a=3$, $r_0/a=0.5$, $q=2$, $s=1$, $\beta=0$, $a/L_n=3$,
  $a/L_{Te}=a/L_{Ti}=a/L_{Tz}=1$, $T_e=T_i=T_z$,
  $\hat{\nu}_{ei}=0$, $\rho_{s0}/a = 0.0035$
\item[] \TEMIII: Electron temperature gradient driven TE
  mode \newline $R_0/a=3$, $r_0/a=0.375$, $q=1.4$, $s=0.8$,
  $\beta=0$, $a/L_{n}=1$, $a/L_{Te}=7/3$, $a/L_{Tz}=7/3$,
  $a/L_{Ti}=0$, $T_e=T_i=T_z$, $\hat{\nu}_{ei}=0$, $\rho_{s0}/a
  = 0.0035$
\end{description}
Here the indices represent electrons ($e$), main ions ($i$) and
impurities ($z$). \red{The density gradient driven case (\TEMII) is one of
the \gyro\ standard cases in the \gyro\ nonlinear gyrokinetic simulation
database \cite{gyrodatabase}. The electron temperature gradient driven
case (\TEMIII) have been used in the fluid simulations presented in
Ref.~\cite{fulopnordman}. } In both cases fully ionized nickel, $Z=28$,
is introduced in trace (i.e. $Z n_z/n_e \ll 1$) quantities
$n_z/n_e=2\times10^{-3}$, however note that $Z^2 n_z/n_e \sim \ord
\left(1\right)$ which is important for the approximate model of the
impurity peaking factor we will use. The use of nickel will ease
comparison with previous work e.g. \cite{puiatti,albert}, but the main
conclusions will be valid for any high-$Z$ impurity.  $R_0$ is the
major radius of the magnetic axis and $r_0$ the local reference minor
radius, $q$ is the safety factor and $s=(r/q)(dq/dr)$ the magnetic
shear, while $\beta$ represents the ratio of plasma pressure to
magnetic pressure. We note that electromagnetic fluctuations appearing
for finite $\beta$ have negligible effect on TE modes as trapped
electrons cannot carry parallel current.  In the main part of the
paper, the plasma is assumed to be hot enough for collisions to be
ignored and consequently the electron-ion collision frequency is
$\hat{\nu}_{ei}=0$, except when it is stated otherwise.

\red{
This paper considers turbulent fluxes. Neoclassical simulations of 
the baseline cases with \neo~\cite{neo} using 
$\hat{\nu}_{ei} = 0.0058 \, c_s/a$ (corresponding to $T_e = 7~\mathrm{keV}$, $n_i = 3 \times 10^{19}~\mathrm{m}^{-3}$, 
$\ln \Lambda = 17$ and $a = 1~\mathrm{m}$) result in fluxes that are 
an order of magnitude smaller than the turbulent fluxes from nonlinear 
\gyro~simulations.
}

\begin{figure}[!ht]
  \scalebox{0.95}{\includegraphics{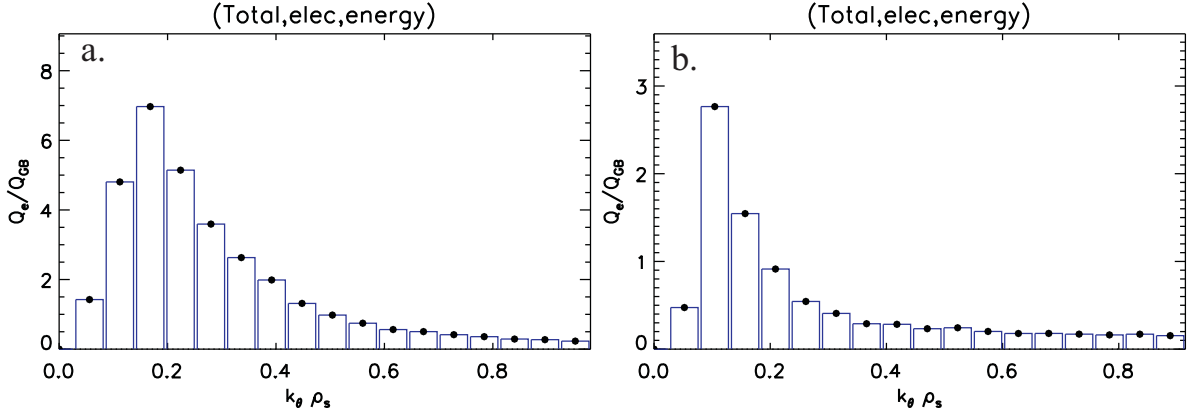}}
  \caption{Normalized electron energy fluxes $Q_e/Q_{\mathrm{GB}}$ as
    functions of poloidal wave-number $k_\theta\rho_s$ from nonlinear
    \gyro~simulations for \TEMII~(a) and \TEMIII~(b).  }
\label{nl}
\end{figure}

\begin{figure}[!ht]
\scalebox{1.0}{\includegraphics{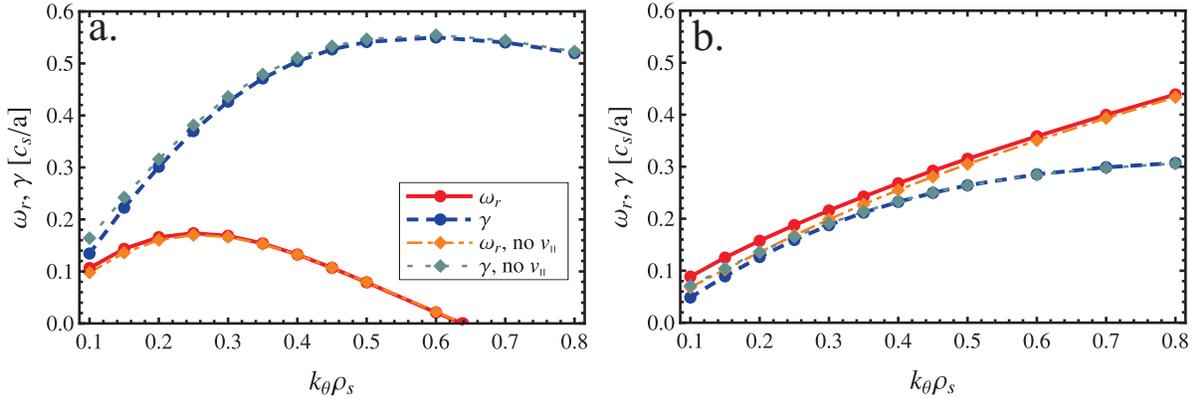}}
\caption{Linear growth rate $\gamma$ (circle markers, blue dashed
  lines) and real mode frequency $\omega_r$ (circle markers, red solid
  lines) as functions of poloidal wave-number $k_\theta\rho_s$ for
  \TEMII~(a) and \TEMIII~(b).  Linear growth rate $\gamma$ (diamond
  markers, green dotted lines) and real mode frequency $\omega_r$
  (diamond markers, orange dash-dotted lines) for the same cases but
  with parallel ion motion neglected in \gyro.  }
\label{freqk}
\end{figure}

Nonlinear \gyro~simulations of the baseline cases show that the
largest fluxes occur in the vicinity of $k_\theta\rho_s=0.15$ for both
of them; see Fig.~\ref{nl} showing the poloidal wave number spectra of
the gyro-Bohm normalized electron energy fluxes.  Here $k_\theta$ is
the poloidal wave-number and $\rho_s = \rho_{s0} \left(1+\epsilon \cos
  \theta\right)$ the ion sound Larmor radius, where $\rho_{s0}$
denotes $\rho_{s}$ at $R_0$, $\epsilon=r_0/R_0$ is the inverse aspect
ratio and $\theta$ the extended poloidal angle.  Consequently
$k_\theta\rho_s=0.15$ is used in the quasilinear simulations for both
cases.  Note, however, that the maximum of the linear growth rates
$\gamma$ are located at higher $k_\theta\rho_s$ as shown in
Fig.~\ref{freqk}.  {As expected for a temperature gradient driven TE
  mode, the frequency increases, i.e. it propagates faster in the
  electron diamagnetic direction, with increasing $k_{\theta}$ in
  \TEMIII~\cite{puiatti}.}  Frequencies are given in $c_s / a$ units,
where $c_s = \left(T_e/m_i\right)^{1/2}$ is the ion sound speed and
$r_0=a$ for the last closed flux surface.

The perturbed electrostatic potential $\phi$ and eigenvalues $\omega =
\omega_r + i \gamma$ are obtained by linear electrostatic gyrokinetic
initial-value calculations with \gyro$\!$~\cite{gyro}. Linear
initial-value studies only consider the most unstable mode and any
sub-dominant modes are neglected. In the simulations a model
Grad-Shafranov magnetic equilibrium was used, where the
$\ord\left(\epsilon\right)$ corrections to the drift frequencies are
retained.  
Flux-tube (periodic) boundary conditions were used, with a 128 point
velocity space grid (8 energies, 8 pitch angles and two signs of
velocity), the number of radial grid points is 6, and the number of
poloidal grid points along particle orbits is 20 for trapped
particles. The location of the highest energy grid point is at $m_i
v^2/\left(2T_i\right)=6$. The ions were taken to be gyrokinetic and
the electrons to be drift kinetic with the mass ratio
$\left(m_i/m_e\right)^{1/2}=60$.

The nonlinear electrostatic \gyro~simulations performed for the
  baseline cases also use gyrokinetic ions and drift kinetic electrons
  and the same velocity resolution as the linear simulations.  At least 18
  toroidal modes are used to model ${1/4}^{\mathrm{th}}$ of the torus,
  with the highest resolved poloidal wave number being
  $k_\theta\rho_s \approx 0.9$.
  The number of radial grid points is 200.  The simulations are run
  with the integration time step $\Delta t = 0.01 a/c_s$ for $t > 200
  a/c_s$.  

\begin{figure}[!ht]
\scalebox{1.05}{\includegraphics{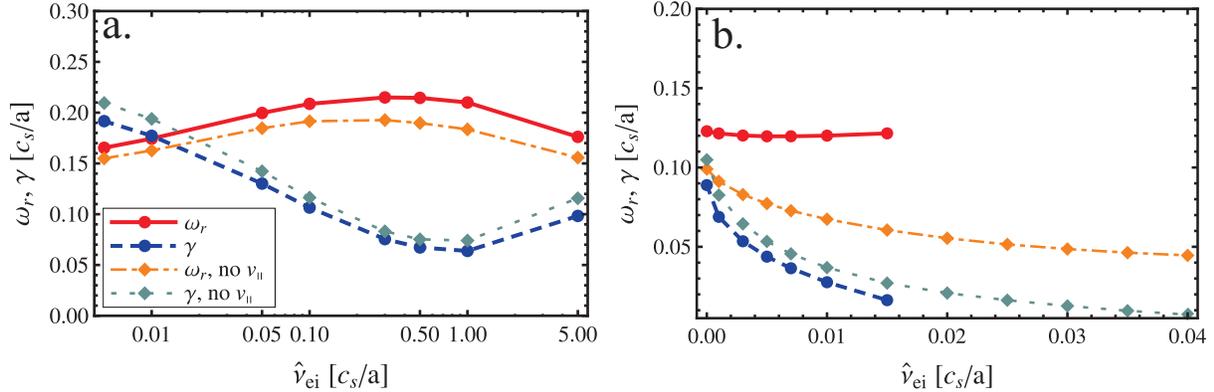}}
\caption{ Linear growth rate $\gamma$ (circle markers, blue dashed
  lines) and real mode frequency $\omega_r$ (circle markers, red solid
  lines) as functions of electron-ion collision frequency
  $\hat{\nu}_{ei}$ for \TEMII~(a) \red{(note the logarithmic $\hat{\nu}_{ei}$-axis)} and \TEMIII~(b).  Linear growth rate
  $\gamma$ (diamond markers, green dotted lines) and real mode
  frequency $\omega_r$ (diamond markers, orange dash-dotted lines) for
  the same cases but with parallel ion motion neglected in \gyro.  
  }
\label{freqnu}
\end{figure}

Introducing a small collision rate is expected to have stabilizing
effect on the collisionless TE mode, because the trapped electrons,
driving the instability, can be detrapped.  One of the most
interesting distinctions between the two different branches of the TE
modes we study concerns the dependence of the linear growth rate on
collisionality. If the TE mode is mainly driven by the electron
temperature gradient, the mode is completely stabilized by collisions
at a very low collision frequency, as was pointed out in
~Ref.~\cite{angcoll}. As shown in Fig.~\ref{freqnu}, this was verified
also in our simulations, where \TEMIII~was suppressed already for
$\hat{\nu}_{ei}>0.015~c_s/a$ while \TEMII~persisted even for 
\red{very high collisionalities, also consistent with earlier studies 
of density gradient driven TE modes \cite{ernst,angcoll}} 
(note that the $\hat{\nu}_{ei}$-ranges
plotted are different, and that $\omega_r$ is positive for modes
propagating to the electron diamagnetic direction according to
\gyro~conventions).  
\red{For $\hat{\nu}_{ei}>1.0~c_s/a$ \TEMII~even exhibits an increase in growth rate with increasing collisionality. 
This could indicate that the mode is turning into a dissipative TE mode, 
but it can also suggest }
that TE modes driven by density gradients 
\red{remain} unstable
even at large collisionalities.  Further it can be noted that
neglecting parallel ion motion in the simulations, leads to a small
\red{to moderate} reduction of the real mode frequency while the
growth rate is almost unaffected.

\begin{figure}[!ht]
\scalebox{1.0}{\includegraphics{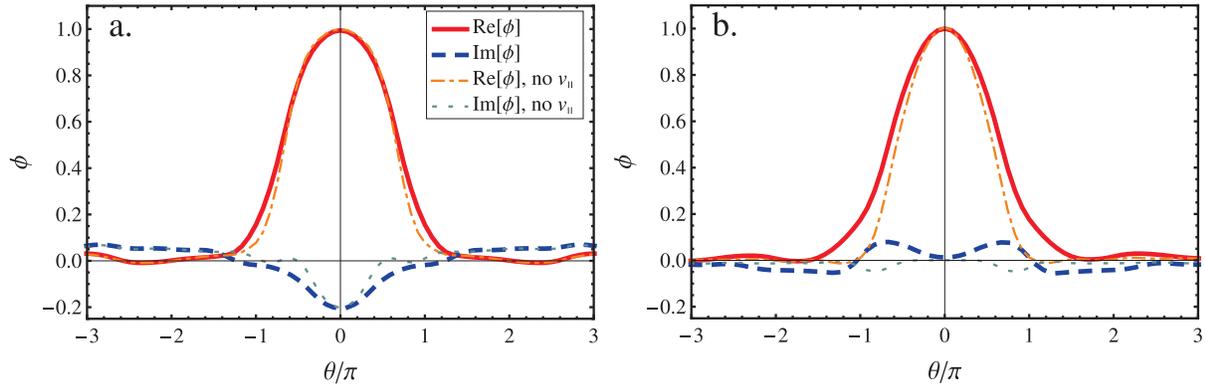}}
\caption{ Linear parallel mode structure of the perturbed potential
  $\phi \left(\theta\right)$ for \TEMII~(a) and \TEMIII~(b).  Real
  part (red solid lines) and imaginary part (blue dashed lines) of
  $\phi$.  Real part (orange dash-dotted lines) and imaginary part
  (green dotted lines) of $\phi$ for the same cases but with parallel
  ion motion neglected in \gyro.  Note that the actual resolution of
  the simulation covers $\theta/\pi = \left(-7, 5\right)$, by
  \gyro~convention.  }
\label{BalloonPhi}
\end{figure}

Figure~\ref{BalloonPhi} shows how the perturbed potential varies with
the extended poloidal angle.  We see that both cases exhibit highly
ballooned structures, concentrated to $\theta \in \left[-\pi,
  \pi\right]$, and that there is no significant difference between
including and not including parallel ion motion (note that the effect
of parallel ion motion is expected to be stronger in cases of lower
$k_\theta\rho_s$).

\section{Impurity density peaking}\label{sec:impurity}
In this section, the zero flux density gradient (peaking factor) for
trace impurities is analyzed. We utilize a semi-analytical model
introduced in Ref.~\cite{varenna}, where the effect of a poloidally
varying equilibrium electrostatic potential $\phi_E$ is included.  The
focus is on the poloidally varying part of the electrostatic
potential, and effects caused by a radial electric field, such as
toroidal rotation, are neglected.  \red{However, we note, that the
  Coriolis drift or the poloidal redistribution of impurities due to
  centrifugal forces can have a non-negligible influence on impurity
  transport, as found in recent works \cite{Casson2010,Camenen2009,angioni2011,angioni2012}.  }
The \red{poloidally} varying potential introduces an $\Ev\times \Bv$
drift frequency labeled $\omega_E$ in the GK equation
(Eq.~(\ref{gke})), which disappears, $\omega_E=0$, in the symmetric
case (note that \gyro~only considers the poloidally symmetric case).
Poloidal variation can be caused by the presence of a species with
strong temperature anisotropy \cite{reinke}, which is the case in
discharges with radio frequency (RF) heating of minority ions on the
outboard side \cite{ingessonppcf,kazakov}.

The poloidally varying potential is assumed to be weak in the sense
that $e \Delta \phi_E/T_{\alpha} \ll 1$ (where $T_{\alpha}$ is the temperature of
species $\alpha$). This implies that the effect of poloidal asymmetries on
the main species can be neglected. This justifies the use of
\gyro~simulations neglecting poloidal asymmetries to obtain linear
mode characteristics. By requiring $Z\gg 1$, we allow $Z e \Delta
\phi_E/T_z \sim \ord \left(1\right)$ and consequently the impurities
can be poloidally asymmetrically distributed. Hence their
$\Ev\times\Bv$ drift in the poloidally varying electrostatic potential
$\phi_E$ is not negligible. This model was presented in
Ref.~\cite{albert} and we refer to this work for more details. The
model we use for the equilibrium electrostatic potential is given by
Eq.~(11) in Ref.~\cite{albert}
\begin{equation}
Ze\phi_E/T_{z}=-\kappa \cos \left(\theta-\delta\right),
\label{eq:Poloidal_potential}
\end{equation}
where $\delta$ represents the angular position where the impurity
density has its maximum and $\kappa$ sets the strength of the poloidal
asymmetry.  Thus the impurity density will be assumed to vary
according to $n_z\left(\theta,r\right) = n_{z0}\left(r\right) \mathcal
N(\theta)$ with $\mathcal N(\theta)=\exp{[\kappa
    \cos{(\theta-\delta)}]}$.  In the model for ion cyclotron
resonance heating (ICRH) driven asymmetries presented in
Ref.~\cite{kazakov} $\delta = \pi$ is obtained, 
\red{
however, since impurity accumulation has also been observed at other
 poloidal locations we shall consider $\delta = 0$ and $\delta = \pi/2$ 
cases as well.}

Note that, in contrast to Ref.~\cite{albert}, the work presented 
here as well as in Ref.~\cite{varenna} retains the effects of the
parallel ion streaming in the GK equation (Eq.~(\ref{gke})).  In the case of
the transport driven by TE mode turbulence, as we will show here, this
term can significantly affect the impurity peaking.

\subsection{Zero flux impurity density gradient}
We consider particle transport driven by a single, representative,
toroidal mode.  The impurity peaking factor is calculated by requiring
the linear impurity flux $\Gamma_z$ to vanish
\begin{equation}
  0 = \left\langle\Gamma_z\right\rangle \equiv \left\langle \Im \left[
  -\frac{k_\theta}{B}\hat{n}_z\phi^\ast \right] \right\rangle
  = \left\langle \Im \left[-\frac{k_\theta}{B}\int d^3v J_0\left(z_z\right) g_z
    \phi^\ast \right] \right\rangle ,\label{ge}
\end{equation} 
where $\langle\cdot \rangle $ denotes the flux surface average,
$\Im\left[ \cdot \right]$ denotes imaginary part, $\hat{n}_z$ is the
perturbed impurity density, $g_z$ the non-adiabatic part of the
perturbed impurity distribution function, $J_0$ is the Bessel function
of the first kind, $z_z=k_\perp v_{\perp}/\omega_{cz}$, $\omega_{cz}=Z
e B/m_z$ is the cyclotron frequency and $k_\perp = \left(1 + s^2
  \theta^2\right)^{1/2} k_\theta$. Furthermore $m_z$ is the impurity
mass, $\phi^\ast$ is the complex conjugate of the perturbed
electrostatic potential, and $B$ is the strength of the equilibrium
magnetic field.  The subscripts $_\|$ and $_\perp$ denote the parallel
and perpendicular directions with respect to the magnetic field.

The non-adiabatic perturbed impurity distribution $g_z$ is obtained
from the linearized GK equation
\begin{equation}
  \left.\frac{v_\parallel}{q R} \frac{\partial {g}_z}{\partial
    \theta}\right|_{\mathcal{E},\mu}    
    -i(\omega- \omega_{Dz}-\omega_E) {g}_z -
  C[g_z] =-i\frac{Z e f_{z0}}{T_z}\left(\omega-\omega_{\ast
    z}^T \right)\phi J_0(z_z),
\label{gke}
\end{equation}
where $\omega = \omega_r + i \gamma$ is the mode frequency,
$f_{z0}=n_{z0}(m_z/2\pi T_z)^{3/2}\exp(-\energy/T_z)$ is the
equilibrium Maxwellian distribution function, $\energy = m_z v^2/2 + Z
e \phi_E$ is the total unperturbed energy, $\mu = m_z \vpe^2/\left(2
  B\right)$ is the magnetic moment, $n_z(\mathbf{r})=n_{z0}\exp[-Z
e\phi_E(\mathbf{r})/T_z]$ is the poloidally varying impurity density
and $n_{z0}$ is a flux function. The diamagnetic frequency is defined
as $\omega_{\ast z}=-k_\theta T_z/Z eB L_{nz}$ and $\omega_{\ast z}^T
=
\omega_{\ast z}\left[1 
  +\left(x^2-3/2\right)L_{nz}/L_{Tz}\right] $, and $x=v/v_{Tz}$
represents velocity normalized to the thermal speed
$v_{Tz}=(2T_z/m_z)^{1/2}$.  The magnetic drift frequency is $
\omega_{Dz}=
-2 k_\theta T_z (x_{\perp}^2/2+x_{\parallel}^2) \mathcal
D\left(\theta\right)/\left(m_z\omega_{cz} R\right) $, where $\mathcal
D\left(\theta\right)= \cos{\theta}+ s \theta \sin{\theta}$.  The
$\Ev\times \Bv$ drift frequency of the particles in the equilibrium
electrostatic field $\omega_E$ is
\begin{equation}
\red{ \omega_{E}=-\frac{k_\theta}{B} \frac{s \theta}{r} \frac{\partial \phi_E}{\partial
    \theta}
}
,
\label{ome}
\end{equation}
and was derived in Appendix A of Ref.~\cite{albert} \red{(here, the
  $\partial \phi_E/\partial r$ part is dropped)}. $C[\cdot]$ is the
collision operator.

A solution to Eq.~(\ref{gke}), and the subsequent expression for the
peaking factor, is presented in Ref.~\cite{varenna}.  It is a
perturbative solution in the small parameter $Z^{-1/2} \ll 1$, keeping
terms up to $\ord(Z^{-1})$ in the expansion of ${g}_z$. This is based
on the fact that $\omega_{Dz}/\omega$, $\omega_{\ast z}^T/\omega$, and
$J_0(z_z)-1\approx -z_z^2/4$ are all $\sim\!1/Z$ small, and that our
ordering $Ze\phi_E/T_z\sim\ord (1)$ requires that $\omega_E/\omega$
also is formally $\sim\!1/Z$ small.  The solution assumes that
impurity self-collisions dominate over collisions with unlike species,
which follows from the ordering $n_z Z^2 / n_e \sim \ord
\left(1\right)$, and the self-collisions are modeled by the full
linearized impurity-impurity collision operator $C_{zz}^{(l)}$,
maintaining the conservation properties and self-adjointness.  As
earlier mentioned it is also assumed that $\phi$ and $\omega$ are
known from the solution of the linear gyrokinetic-Maxwell system
(obtained from \gyro) and that they are unaffected by the presence of
trace impurities, and in particular their poloidal asymmetry. The
impurity transit frequency $v_\parallel/\left(q R\right)$ is typically
much smaller than the mode frequency $\omega$, and therefore magnetic
(and electrostatic) trapping of the impurities can be neglected.

The expression for the impurity peaking factor, $a / L_{nz}^0$, is
given in Eq.~(8) of Ref.~\cite{varenna} and can be modified into
\begin{equation}
\frac{a}{L_{nz}^0}= 2 \frac{a}{R_0}\langle \mathcal{D} \rangle_\phi +
\frac{a}{r}s\kappa\langle\theta \sin(\theta-\delta) \rangle_\phi -
\frac{2 a^2}{(q R_0)^2 k_{\theta} \rho_{s0}}\frac{Z m_i}{m_z}
\frac{c_s}{a} \frac{\omega_r}{\omega_r^2+\gamma^2} \left\langle \left|
\frac{\partial \phi}{\partial \theta}\right|^2/
\left|\phi\right|^2\right\rangle_\phi,
\label{PeakingFactor}
\end{equation}
where $\langle\dots\rangle_\phi=\langle\dots \mN
|\phi|^2\rangle/\langle\mN |\phi|^2\rangle$.  To find this expression
$\ord(\epsilon)$ corrections together with finite values of the mode
eigenfunction outside the range $[-\pi,\pi]$ of the extended poloidal
angle have been neglected.  As a consequence the expression is not
valid in cases of highly elongated ballooning eigenfunctions, but as
shown in Fig.~\ref{BalloonPhi} the TE modes we study have a $\phi$
localized to this interval.  It is interesting to note that up to the
considered order, $\ord(Z^{-1})$, both finite Larmor radius (FLR)
effects and the effects of collisions do not appear.  Furthermore we
see that $a / L_{nz}^0$ consists of three terms: the first term of
Eq.~(\ref{PeakingFactor}) represents the contribution of the magnetic
drift, the second term stems from the $\Ev\times \Bv$ drift and is
only non-zero when there is a poloidally varying potential, and the
last term arises because of the impurity parallel dynamics. The first
two terms were present already in Eq.~(14) of Ref.~\cite{albert}, but
in that expression parallel ion/impurity dynamics was neglected.  The
term due to parallel dynamics contains only non-negative quantities,
except $\omega_r$, and consequently impurity parallel dynamics acts to
increase (decrease) the impurity peaking if $\omega_r$ is negative
(positive). This leads to the conclusion that when the TE mode is the
dominant instability ($\omega_r > 0$) the parallel dynamics should act
to decrease the impurity peaking, while the opposite is true for ITG
modes. \red{We note that in the poloidally symmetric ($\mN=1$) case the parallel
compressibility term in Eq.~(\ref{PeakingFactor}) is consistent with
Eqs.~(9) and (10) of Ref.~\cite{angioni2006} in the high $Z$ limit when $k_\|^2$
is defined as
$(qR_0)^{-2}\langle|\partial_{\theta}\phi|^2\rangle/\langle|\phi|^2\rangle$.} {The
  effect of parallel dynamics was pointed out already in
  \cite{angioni2006}, where it was observed that in conditions of
  strong electron temperature gradient the mechanism can be large
  enough to reverse the total pinch of trace impurities from inwards
  to outwards.}

\subsection{Parametric dependences in poloidally symmetric cases}
\label{SymmetricSection}

In the present section we show how the impurity peaking factor and
mode eigenvalues depend on electron temperature gradient,
ion-to-electron temperature ratio, electron density gradient and
safety factor, when $\phi_E$ is poloidally symmetric.  Results are
presented for both \TEMII~and \TEMIII, when parallel ion/impurity
dynamics are included as well as when they are neglected. Results from
Eq.~(\ref{PeakingFactor}) are compared to simulations by \gyro. 
  Although the analytical model for the peaking factor given in
  Eq.~(\ref{PeakingFactor}) is based on a single linear mode, for
  completeness we will present a few cases where peaking factors have
  been determined from nonlinear simulations with \gyro.  The reason
  for this is to provide further insight into the validity of the
  analytical model.  
  Note, that the analytical results from
Eq.~(\ref{PeakingFactor}) are not expected to agree exactly even with
the linear results of \gyro\, since the approximation only retains
effects up to order $1/Z$ as well as neglects $\ord(\epsilon)$
corrections.  Note that for cases with parallel ion/impurity dynamics
neglected, the mode characteristics come from \gyro~simulations where
parallel compressibility effects have been turned off. Therefore they
differ slightly from the magnetic drift contribution in cases where
parallel dynamics is included. This can be observed in
e.g. Fig.~\ref{aLnz0aLTe}a-b, comparing the blue dotted line with the
orange dashed line.

\subsubsection{Temperature and temperature gradient dependences}
\begin{figure}[htbp]
  \scalebox{1.0}{\includegraphics{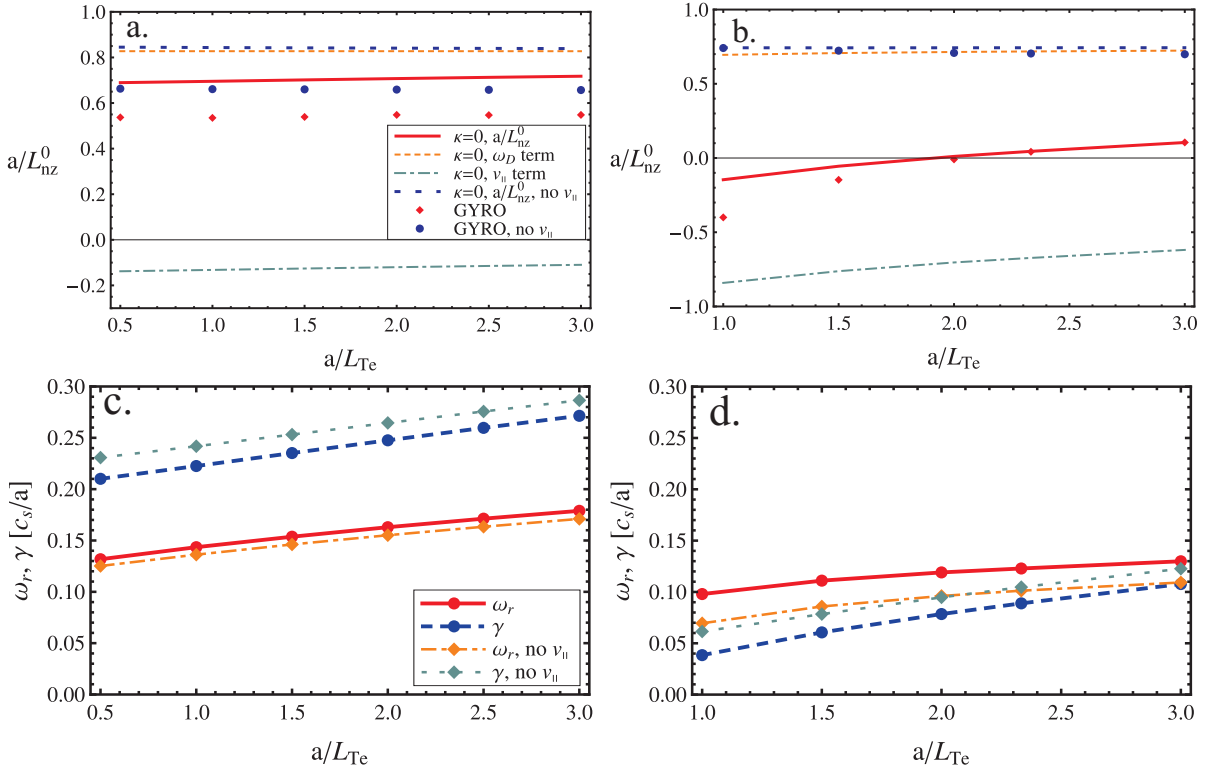}}
  \caption{(a, b)~Impurity peaking factor for trace nickel as function of
  electron temperature gradient $a/L_{Te}$ for \TEMII~(a) and \TEMIII~(b). Red solid line is
  the peaking factor from Eq.~(\ref{PeakingFactor}), orange dashed line
  the magnetic drifts contribution, and green dash-dotted line the
  parallel compressibility contribution. Blue dotted line is the
  peaking factor from Eq.~(\ref{PeakingFactor}) without parallel
  compressibility effects. Red diamonds and blue dots correspond to
  \gyro~results with and without parallel compressibility effects
  respectively.\newline
    (c, d)~Linear growth rate $\gamma$ (circle markers, blue dashed lines) and real mode frequency $\omega_r$ (circle markers, red solid lines) as functions of electron temperature gradient $a/L_{Te}$ for \TEMII~(c) and \TEMIII~(d). 
    Linear growth rate $\gamma$ (diamond markers, green dotted lines) and real mode frequency $\omega_r$ (diamond markers, orange dash-dotted lines) for the same cases but with parallel ion motion neglected in \gyro.
    }
    \label{aLnz0aLTe}
\end{figure}

Figure~\ref{aLnz0aLTe}a-b shows how the nickel peaking factor varies
with electron temperature gradient and Fig.~\ref{aLnz0aLTe}c-d the
corresponding eigenvalues.  For both cases there is a slight increase
in peaking for increasing $a/L_{Te}$, but the dependence is generally
very weak.  This is what is expected from Eq.~(\ref{PeakingFactor}),
where there is no explicit dependence on $a/L_{Te}$.  Instead the
variations are caused by changes in mode frequency, shown in
Fig.~\ref{aLnz0aLTe}c-d, and perturbed potential through the parallel
compressibility term, which is reflected in the fact that almost no
variation at all is observed for the results with parallel
compressibility neglected.  Both the linear growth rate and real mode
frequency increase with electron temperature gradient, where the
increase in growth rate is expected since electron temperature
gradient is one of the drives for TE modes.
  
\begin{figure}[!ht]
	\scalebox{1.0}{\includegraphics{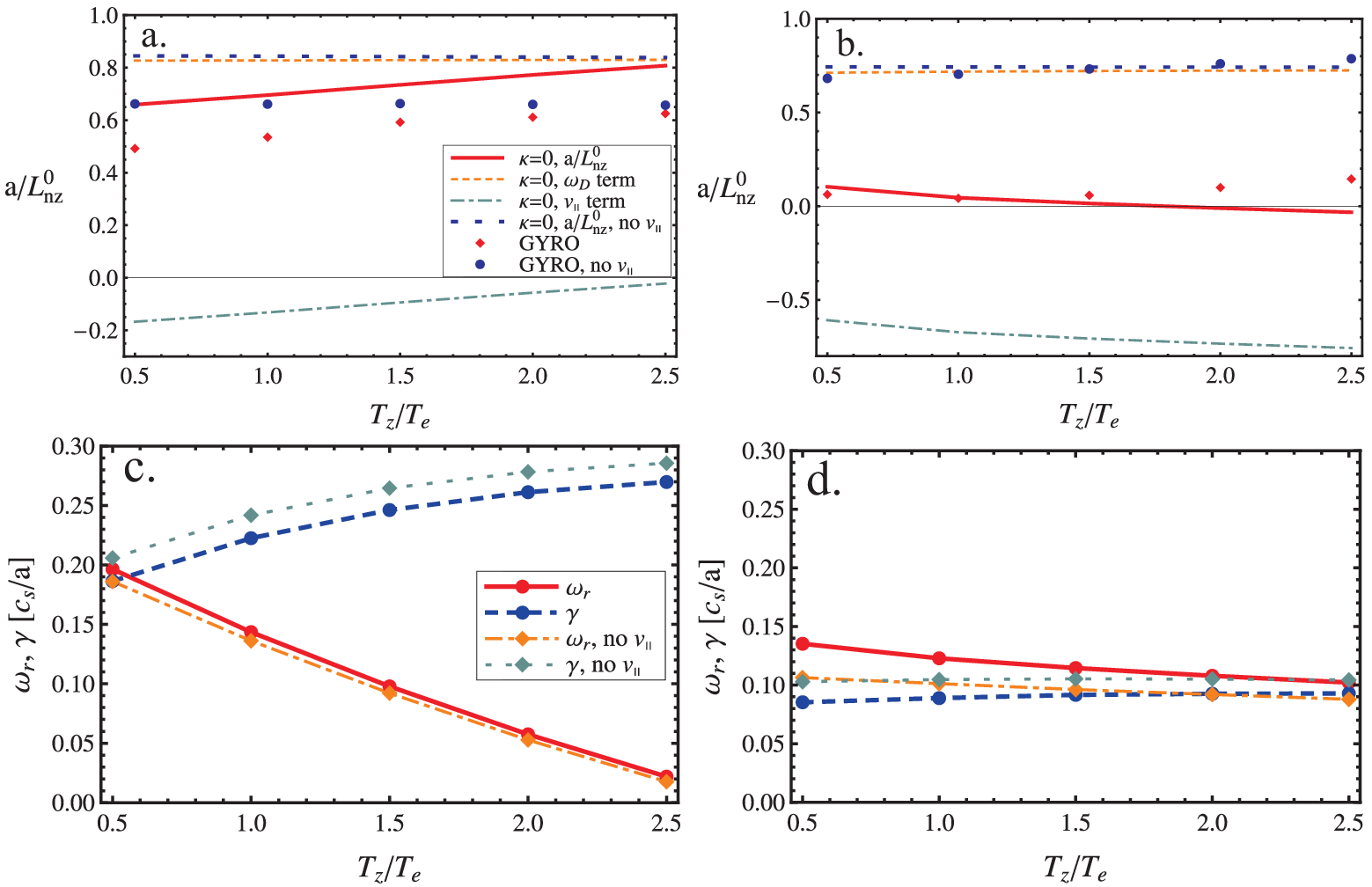}}
	\caption{
	(a, b)~Impurity peaking factor for trace nickel as function of
  ion-to-electron temperature ratio $T_i/T_e$ (note that $T_i/T_e = T_z/T_e$) 
  for \TEMII~(a) and \TEMIII~(b). Red solid line is
  the peaking factor from Eq.~(\ref{PeakingFactor}), orange dashed line
  the magnetic drifts contribution, and green dash-dotted line the
  parallel compressibility contribution. Blue dotted line is the
  peaking factor from Eq.~(\ref{PeakingFactor}) without parallel
  compressibility effects. Red diamonds and blue dots correspond to
  \gyro~results with and without parallel compressibility effects
  respectively.\newline
    (c, d)~Linear growth rate $\gamma$ (circle markers, blue dashed lines) and real mode frequency $\omega_r$ (circle markers, red solid lines) as functions of ion-to-electron temperature ratio $T_i/T_e$ for \TEMII~(c) and \TEMIII~(d). 
    Linear growth rate $\gamma$ (diamond markers, green dotted lines) and real mode frequency $\omega_r$ (diamond markers, orange dash-dotted lines) for the same cases but with parallel ion motion neglected in \gyro.
    }
    \label{aLnz0TiTe}
\end{figure}
The peaking factor dependence on ion-to-electron temperature ratio
(note that $T_i/T_e = T_z/T_e$) is illustrated in
Fig.~\ref{aLnz0TiTe}a-b. We see that it exhibits a similar dependence
as to $a/L_{Te}$, which is weak and only enters through changes in
$\omega_r$, $\gamma$ and $\phi$.  How $\omega_r$ and $\gamma$ are
affected by changes in $T_i/T_e$ is shown in Fig.~\ref{aLnz0TiTe}c-d.
\TEMII~shows a significant increase in growth rate with $T_i/T_e$ and
decrease in real mode frequency, while \TEMIII~is almost unaffected.
This implies that, as $T_i/T_e$ increases, \TEMII~is approaching an
ITG mode. 
Similar trends were reported in terms of mode frequencies and
growth rates in Ref.~\cite{pusztai} where temperature and density
gradient driven TE mode dominated plasmas were compared (having Ohmic
and electron cyclotron heating, respectively); in the density gradient
driven case $\omega_r$ ($\gamma$) was found to decrease (increase) with
increasing $T_i/T_e$, while weaker temperature ratio dependences of
$\omega_r$ and $\gamma$ were observed in the temperature gradient
driven case.

Earlier studies of TE modes have shown that the linear growth rate is
expected to decrease with increasing $T_i/T_e$, as seen in Fig.~5 of
Ref.~\cite{lang}, which is seemingly in disagreement with what is
observed here. However, it has to be noted that the simulation in
Ref.~\cite{lang} was performed keeping $k_{\theta} \rho_i$ fixed,
while here $k_{\theta} \rho_s$ is fixed. Since $k_{\theta} \rho_s \sim
\left(T_e/T_i\right)^{1/2} k_{\theta} \rho_i$, a parametric scan over
$T_i/T_e$ keeping $k_{\theta} \rho_i$ fixed, results in varying
$k_{\theta} \rho_s$ accordingly and thus these scalings are not
comparable.  
\red{
Furthermore in Ref.~\cite{lang} results are presented with $\gamma$ 
normalized to $v_{Ti}/L_n \propto \sqrt{T_i}$, while here we normalize 
to $c_s / a \propto \sqrt{T_e}$.} 
A test was performed, varying $k_{\theta} \rho_s$ to keep
$k_{\theta} \rho_i$ fixed, and a similar decrease with increasing
$T_i/T_e$ was found.
\red{Also
Ref.~\cite{Peeters2005} finds an increase in linear growth rate 
with increasing $T_i/T_e$ in agreement with our observations.
}
  
A similar rather weak dependence of the impurity peaking on $a/L_{Te}$
and $T_i/T_e$ was found in gyrokinetic simulations by \cite{puiatti},
as long as the most unstable mode remained the same.  As observed in
\TEMII~here, for density gradient driven TE modes the peaking factor
is typically positive.  For temperature gradient driven TE modes
however it can be negative which is also found in \TEMIII~where it is
close to zero or even below, as seen in Figs.~\ref{aLnz0aLTe}b and
\ref{aLnz0TiTe}b.

There is a small discrepancy between the values found by \gyro~and the
values calculated from Eq.~(\ref{PeakingFactor}).  For
\TEMII~Eq.~(\ref{PeakingFactor}) systematically overestimates the
magnetic drift contribution, while for \TEMIII~discrepancies arise
mainly due to the parallel compressibility term.  Still we see that
the effect of the parallel impurity dynamics is rather well explained
by the approximate solution.

From Figs.~\ref{aLnz0aLTe}--\ref{aLnz0TiTe}c-d we see that neglecting
the parallel ion motion in the GK equation ($v_\parallel/\left(q
  R\right) \partial {g}_z/\partial\theta = 0$ in Eq.~(\ref{gke})) only
has a minor impact on the TE mode frequencies.

\subsubsection{Density gradient dependence}
\begin{figure}[htbp]
\scalebox{1.0}{\includegraphics{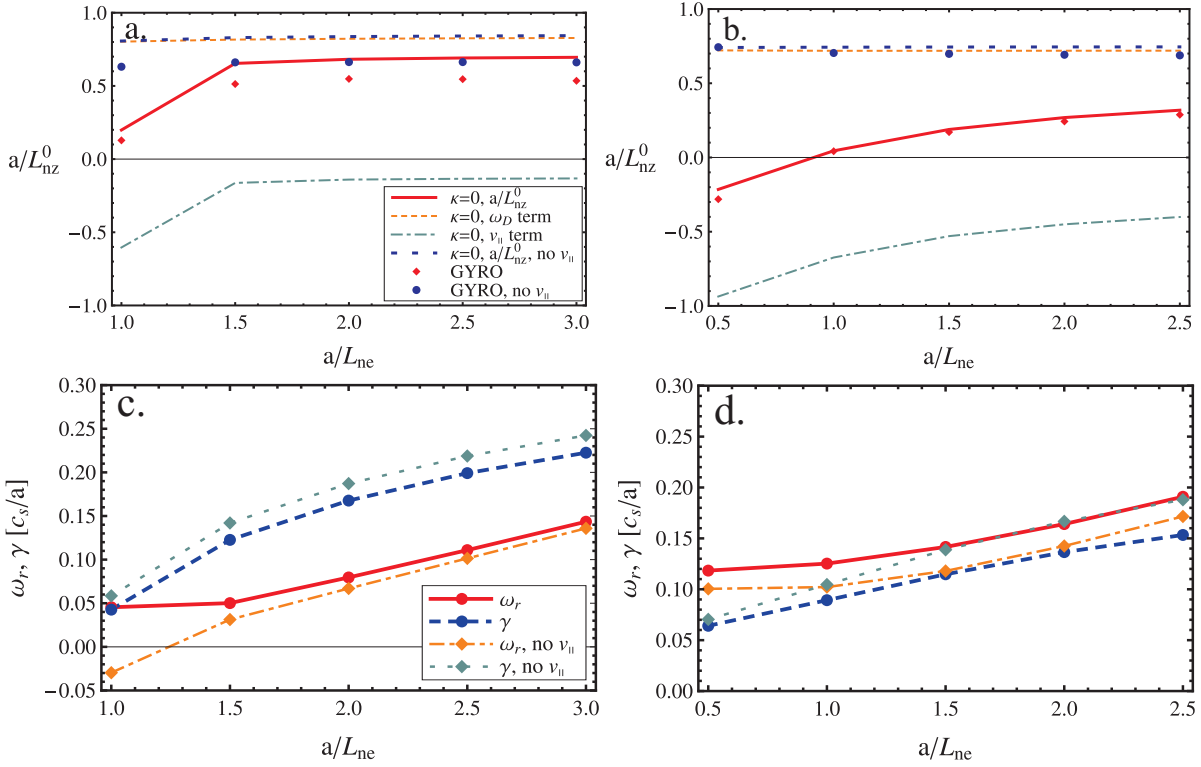}} 
\caption{(a, b)~Impurity peaking factor for trace nickel as function of
  electron density gradient $a/L_{ne}$ for \TEMII~(a) and \TEMIII~(b). Red solid line is
  the peaking factor from Eq.~(\ref{PeakingFactor}), orange dashed line
  the magnetic drifts contribution, and green dash-dotted line the
  parallel compressibility contribution. Blue dotted line is the
  peaking factor from Eq.~(\ref{PeakingFactor}) without parallel
  compressibility effects. Red diamonds and blue dots correspond to
  \gyro~results with and without parallel compressibility effects
  respectively.\newline
    (c, d)~Linear growth rate $\gamma$ (circle markers, blue dashed lines) and real mode frequency $\omega_r$ (circle markers, red solid lines) as functions of electron density gradient $a/L_{ne}$ for \TEMII~(c) and \TEMIII~(d). 
    Linear growth rate $\gamma$ (diamond markers, green dotted lines) and real mode frequency $\omega_r$ (diamond markers, orange dash-dotted lines) for the same cases but with parallel ion motion neglected in \gyro.
}
\label{aLnz0aLne}
\end{figure}
The scaling of the impurity density peaking with the main species
density peaking is shown in Fig.~\ref{aLnz0aLne}a-b, and the
corresponding eigenvalues in Fig.~\ref{aLnz0aLne}c-d.  It is
interesting to note how the impurity peaking shows a significant
decrease in \TEMII~for $a/L_{ne} = 1.0$, compared to the other values,
which is almost solely due to the change in the factor
$\omega_r/\left(\omega_r^2+\gamma^2\right)$ in the parallel
compressibility term of Eq.~(\ref{PeakingFactor}) (the variation in
$\phi \left(\theta\right)$ with $a/L_{ne}$ is rather weak).  The
change in magnitude of this factor is clearly observed in
Fig.~\ref{aLnz0aLne}c, where for $a/L_{ne} = 1.0$\red{,} $\omega_r$ and
$\gamma$ are comparable in size while for the other points $\gamma$ is
significantly larger. 
For density gradient driven TE modes a strong
reduction in the peaking factor towards weaker density gradients has
previously been reported in Ref.~\cite{skyman}, using both quasilinear
and nonlinear {\sc gene} \cite{gene} simulations. 
In
\TEMIII~$\omega_r/\left(\omega_r^2+\gamma^2\right)$ experience a small
reduction with increasing $a/L_{ne}$, and this is reflected in the
impurity peaking factor which is increased because of the smaller size
of the parallel compressibility term.  For $a/L_{ne} = 0.5$ this case
has a negative peaking factor.

For $a/L_{ne} = 1.0$ we could expect that \TEMII~changes to become a
temperature gradient driven TE mode, since then the density and
temperature gradients are comparable ($a/L_{Te} = 1.0$ for \TEMII). It
seems that for temperature gradient driven TE modes the magnitude of
the ratio of $\omega_r$ compared to $\gamma$ is usually larger than
for density gradient driven TE modes. This has the consequence that
the parallel dynamics becomes more important in reducing the impurity
density peaking for temperature gradient driven TE modes.  Further, we
note that as could be expected from Eq.~(\ref{PeakingFactor}) the
peaking factors are unaffected by changes in $a/L_{ne}$ if parallel
compressibility is neglected.  For both cases the linear growth rate
and real mode frequency increase with increasing $a/L_{ne}$.  The
eigenvalues do not change much by neglecting parallel compressibility,
except in \TEMII~with $a/L_{ne} = 1.0$ where an ITG mode is found
instead of a TE mode.

\subsubsection{Safety factor dependence}
\begin{figure}[htbp]
\scalebox{1.0}{\includegraphics{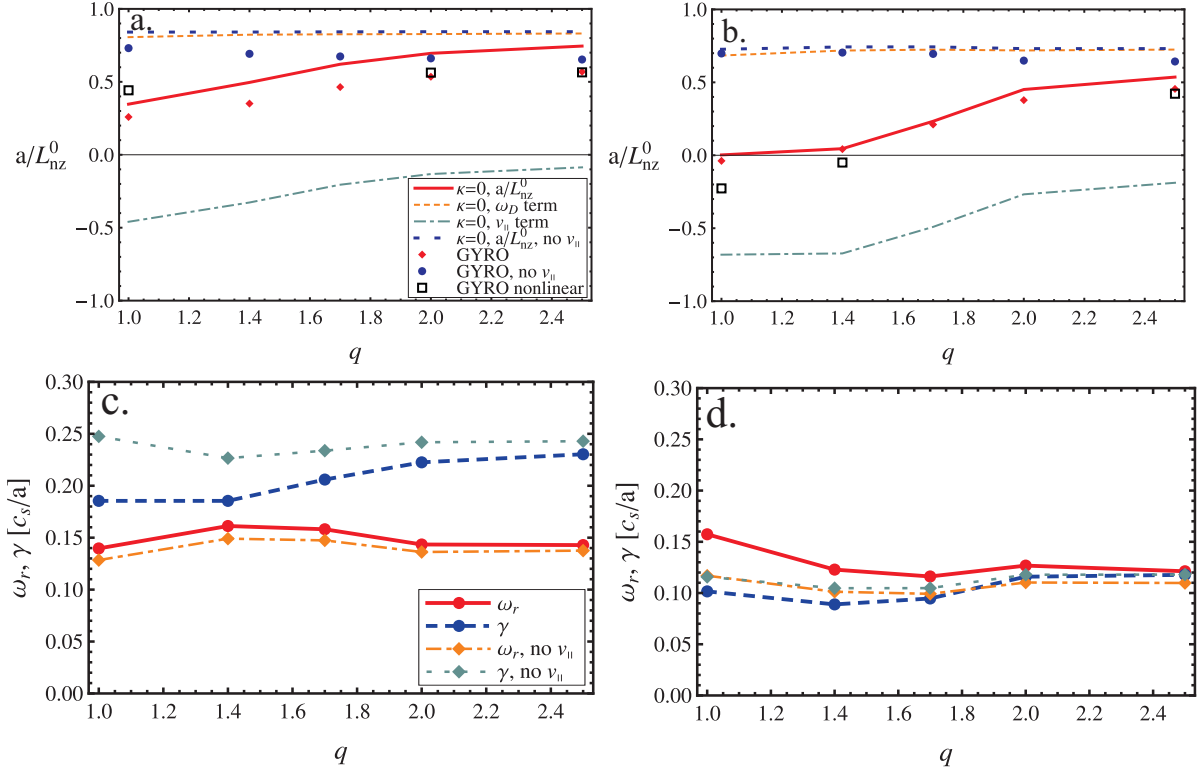}} 
\caption{(a, b)~Impurity peaking factor for trace nickel as function of
  safety factor $q$ for \TEMII~(a) and \TEMIII~(b). Red solid line is
  the peaking factor from Eq.~(\ref{PeakingFactor}), orange dashed line
  the magnetic drifts contribution, and green dash-dotted line the
  parallel compressibility contribution. Blue dotted line is the
  peaking factor from Eq.~(\ref{PeakingFactor}) without parallel
  compressibility effects. Red diamonds and blue dots correspond to
  \gyro~results with and without parallel compressibility effects
  respectively,
  while black hollow squares are results from nonlinear \gyro~runs.
  \newline
    (c, d)~Linear growth rate $\gamma$ (circle markers, blue dashed lines) and real mode frequency $\omega_r$ (circle markers, red solid lines) as functions of safety factor $q$ for \TEMII~(c) and \TEMIII~(d). 
    Linear growth rate $\gamma$ (diamond markers, green dotted lines) and real mode frequency $\omega_r$ (diamond markers, orange dash-dotted lines) for the same cases but with parallel ion motion neglected in \gyro.
    }
\label{aLnz0q}
\end{figure}
From the last term of Eq.~(\ref{PeakingFactor}), it is expected that
the influence of impurity parallel dynamics on impurity peaking factor
is strongly reduced with increasing safety factor. This is also what
is observed in simulations, where the parallel compressibility
contribution is significantly more negative with a small safety factor
and as a consequence the peaking factor is more strongly reduced (see
Fig.~\ref{aLnz0q}a-b).  Since the safety factor typically is higher
closer to the edge, it could be expected that in TE mode dominated
plasmas the impurity parallel dynamics will reduce the peaking factor
more and more, the closer to the core we look, while the opposite
effect is expected in ITG dominated plasmas, \red{as is confirmed by
  simulations presented in Ref.~\cite{Howard2012}.}  However, as
discussed in \cite{dannertjenko}, \red{in a study of collisionless TE
  modes,} it should be noted that the $k_{\theta}$ leading to the
largest fluxes is approximately inversely proportional to $q$, which
is an effect we miss by keeping $k_{\theta}$ constant in our linear
\gyro~simulations. Since in the parallel compressibility contribution
of Eq.~(\ref{PeakingFactor}) there is a factor $1/\left(q^2
  k_{\theta}\right)$, it is reduced to $1/q$ if the mode leading to
the largest fluxes should be considered.  Furthermore it can also be
noted that although the mode frequencies are relatively independent of
$q$ (as shown in Fig.~\ref{aLnz0q}c-d), they are not independent of
$k_{\theta} \rho_s$ (see Fig.~\ref{freqk}) but can on a very crude
estimate be expected to vary linearly with $k_{\theta} \rho_s$ around
the range of $k_{\theta} \rho_s$ we analyze. Because of the factor
$\omega_r/\left(\omega_r^2+\gamma^2\right)$ also found in the parallel
compressibility contribution, this would imply that the dependence on
$q$ is completely canceled for the linear analysis of the mode leading
to the largest fluxes.

Figure~\ref{aLnz0q}a-b also includes impurity peaking factors
determined from nonlinear \gyro~simulations. These were calculated by
linear interpolation of the impurity fluxes from two nonlinear
\gyro~simulations with different impurity density gradient.  In these
simulations, to keep an optimal $k_\theta\rho_s$-resolution around the
peak part of the nonlinear energy and particle flux spectra, the
spacing between the simulated toroidal mode numbers are changed from
case to case while the total number of toroidal modes are held fixed.
\red{In the nonlinear simulation for the density gradient driven case
  we see a significantly weaker, but still existent, $q$-dependence
  than what the fixed-$k_\theta$ linear simulation predicts. This may
  be understood from the above reasoning about the shift in the peak
  of the turbulent spectrum. On the other hand, in \TEMIII~the trend
  is found to be similar to the linear predictions, and interestingly,
  the $q$-dependence is even stronger in the nonlinear case.  }

The difference between including and not-including parallel ion
dynamics for lower values of $q$ is more pronounced 
in the electron
temperature gradient driven \TEMIII~than in the density gradient
driven \TEMII~(a trend, consistent with the results of
Ref.~\cite{pusztai}).
This is mostly due to that the factor
$\omega_r/\left(\omega_r^2+\gamma^2\right)$ in
Eq.~(\ref{PeakingFactor}) is larger for \TEMIII~than for \TEMII~which
can be seen from Fig.~\ref{aLnz0q}c-d.  Accordingly we could expect
that for modes where this factor is relatively large, the effect of
parallel impurity motion on the impurity peaking factor is
strengthened.  The differences in the dependence on $q$ can only come
through the $q$-dependence of $\omega$ and $\phi\left(\theta\right)$,
since all the other parameters appearing in the third term of
Eq.~\ref{PeakingFactor} are the same in the two cases for a fixed $q$.

\subsection{Poloidally asymmetric case}
\begin{figure}[htbp]
\begin{center}
\scalebox{1.0}{\includegraphics{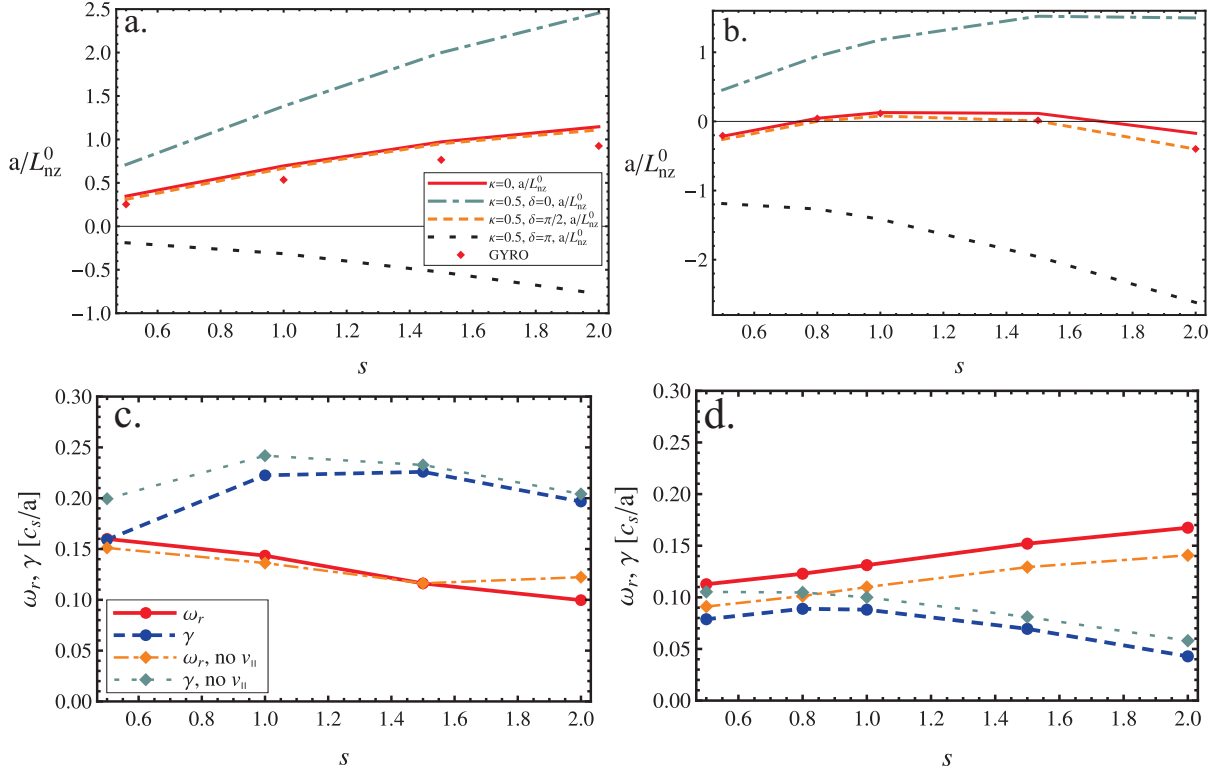}}
  \caption{(a, b)~Impurity peaking factor for trace nickel as function of magnetic shear $s$ for \TEMII~(a) and \TEMIII~(b). 
  \red{Red solid line is the peaking factor from Eq.~(\ref{PeakingFactor}) in the symmetric case, green dash-dotted line corresponds to out-in asymmetry, orange dashed line corresponds to up-down asymmetry, and black dotted line corresponds to in-out asymmetry.}
  Red diamonds correspond to \gyro~results.\newline
  (c, d)~Linear growth rate $\gamma$ (circle markers, blue dashed lines) and real mode frequency $\omega_r$ (circle markers, red solid lines) as functions of magnetic shear $s$ for \TEMII~(c) and \TEMIII~(d). 
  Linear growth rate $\gamma$ (diamond markers, green dotted lines) and real mode frequency $\omega_r$ (diamond markers, orange dash-dotted lines) for the same cases but with parallel ion motion neglected in \gyro.
}
\label{Fig:peakingShear}
\end{center}
\end{figure}
In earlier studies with a poloidally varying potential present,
magnetic shear has been emphasized as one of the most important
parameters affecting the impurity peaking \cite{albert, varenna}. This
can be understood from its explicit appearance in the $\Ev\times \Bv$
drift term of Eq.~(\ref{PeakingFactor}). Consequently this section
will focus on how the impurity peaking factor varies with magnetic
shear, and results will be presented for the poloidally asymmetric
\red{cases} with $\kappa = 0.5$ and 
\red{
$\delta = 0$ (out-in asymmetry), $\delta = \pi/2$ (up-down asymmetry) and
$\delta = \pi$ (in-out asymmetry)}
in Eq.~(\ref{eq:Poloidal_potential}).  We will omit the parametric
dependence on other parameters in this section, because of the
structure of the $\Ev\times \Bv$ drift term of
Eq.~(\ref{PeakingFactor}) and refer to the results of
Sec.~\ref{SymmetricSection}. The inclusion of the $\Ev\times \Bv$
drift term typically leads to the addition of a 
constant to
the peaking factor, for other scalings than with $s$. This is because
the only way this term can change in these scalings is through
$|\phi|^2$, and it is not varying by much.

Figure~\ref{Fig:peakingShear}a-b shows how the peaking factor depends
on magnetic shear for \TEMII~and \TEMIII~in both the symmetric and
asymmetric cases. For \TEMII~the symmetric peaking factor is mainly
governed by the contribution from the magnetic drifts, which increases
with $s$. The contribution from parallel compressibility is relatively
small, and it is not affected much by a change in $s$.  When
introducing the asymmetry, the peaking factor 
\red{changes} significantly
because of the $\Ev\times \Bv$ drift term which then is 
non-zero. 
\red{
In the in-out asymmetric case there is a strong decrease of the impurity peaking, 
while on the contrary in the out-in asymmetric case there is a strong increase. 
The peaking factor remains almost unaffected by an up-down asymmetry.}
Since 
\red{the $\Ev\times \Bv$ drift} 
term becomes larger in magnitude with increasing
$s$, the difference between the symmetric and asymmetric peaking
factor is also increased with $s$. 
The reason why the $\Ev\times \Bv$
drift term leads to a reduction \red{(an increase)} 
of the peaking factor 
\red{for inboard (outboard) impurity accumulation}
is because that
for $\delta = \pi$ \red{($\delta = 0$) the term} 
$\langle\theta
\sin(\theta-\delta) \rangle_\phi$ is negative \red{(positive)}. 
Note however that if 
$s < 0$, this term 
\red{changes sign and leads to an increase (decrease) 
for inboard (outboard) impurity accumulation,}
as shown in
\cite{albert}.  The peaking factor in \TEMIII~shows a similar behavior
to \TEMII, with the main difference being that the contribution from
the parallel compressibility term is significantly larger. Because of
this term, the peaking factor can be negative even in the symmetric
case.

Furthermore in Fig.~\ref{Fig:peakingShear}c-d it can be noted that the
linear eigenvalues are not strongly affected by changes in magnetic
shear, although for $s \geq 1.0$ there is a small stabilizing effect
with increasing shear. Ref.~\cite{lang} reported on a stabilizing
effect which was weaker if the density gradients were large. This is
consistent with what is observed here, since \TEMII~has a larger
density gradient than \TEMIII~ and the stabilizing effect is weaker in
\TEMII.  An increasing positive magnetic shear can stabilize a TE mode
through FLR effects, but it can also drive the mode more unstable by
increasing bad magnetic curvature.  The dependence of linear growth
rate on shear is consequently not trivial.  The eigenvalues do not
change much if the parallel ion dynamics is neglected.

\subsection{Collisions}
\label{sec:collisions}
\red{ The model represented by Eq.~(\ref{PeakingFactor}) models
  impurity self-collisions by the full linearized impurity-impurity
  collision operator $C_{zz}^{(l)}$, and it is found that up to the
  considered order, $\ord(Z^{-1})$, the effect of collisions does not
  appear explicitly.  It thus predicts that the only way for
  collisions to affect the impurity peaking, is through their impact
  on the mode characteristics. This leads to a conclusion that in
  reality collisions should have a relatively weak influence on the
  impurity peaking factor.
  As many of
  the easily accessible gyrokinetic tools employ
  non-momentum-conserving model operators, it is interesting to see
  whether or not the form of the collision operator affects the above
  result. }

\begin{figure}[htbp]
\begin{center}
\scalebox{1.0}{\includegraphics{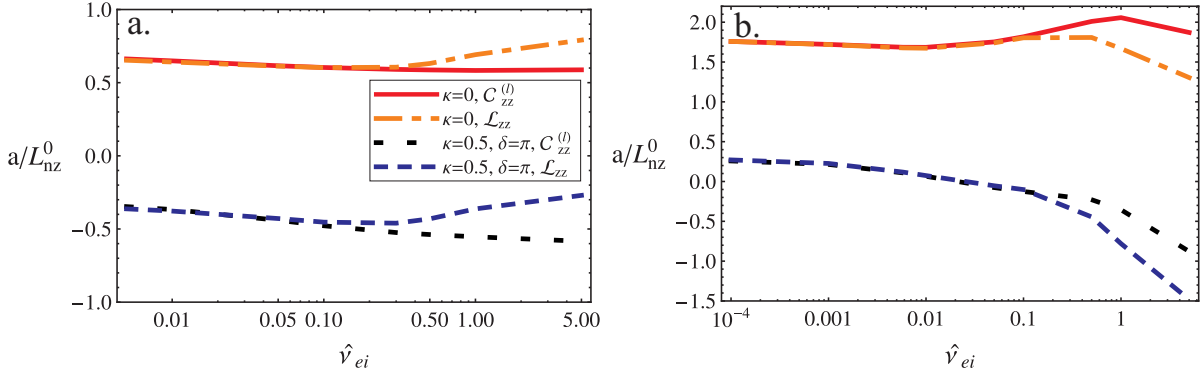}}
  \caption{
  \red{
  Impurity peaking factor for trace nickel as function of electron-ion collision frequency $\hat{\nu}_{ei}$ for \TEMII~(a) and an ITG dominated case (b) (note the logarithmic $\hat{\nu}_{ei}$-axis). 
  Red solid line is the peaking factor from Eq.~(\ref{PeakingFactor}) in the symmetric case and black dotted line the corresponding in the in-out asymmetric case.
  Orange dash-dotted line is the peaking factor in the symmetric case from a model that utilize the Lorentz collision operator, 
  and blue dashed line the corresponding in the in-out asymmetric case.
  }
}
\label{Fig:CollisionComparison}
\end{center}
\end{figure}

\red{In Appendix A we present an alternative model which uses the
  Lorentz (or ``pitch-angle scattering'') operator
  to model
  impurity self-collisions, but is otherwise similar to the
  perturbative solution represented by Eq.~(\ref{PeakingFactor}).  The
  most striking difference between the two models, is the appearance
  of the factor $1 / \left(1 + i \nu_D\left(x\right) / \omega\right)$
  in the contribution related to parallel dynamics, where
  $\nu_D\left(x\right) \; \left[\propto \hat{\nu}_{ei}\right]$ is the
  deflection frequency. This implies that in the case of the Lorentz
  operator the effect of collisions appears explicitly in the
  expression for the impurity peaking factor, which is an artifact of
  $C_{zz}[\vpa f_{z0}]$ being different from
  zero. Figure~\ref{Fig:CollisionComparison} shows a comparison
  between the two models for \TEMII, but also for an ITG dominated case 
  (earlier studied in \cite{varenna} also using fully ionized trace Nickel 
  with local profile and geometry parameters: $r/a=0.3$, $R_0/a=3$, 
  $k_\theta\rho_s=0.3$, $q=1.7$,  $s=1.5$, $a/L_{ne}=1.5$, $T_i/T_e=0.85$, 
  $a/L_{Te}=2$ and $a/L_{Ti}=2.5$).
  \TEMIII~is not considered since it is stabilized already at 
  low collisionality as shown in Fig.~\ref{freqnu}.
  The scalings illustrate that the impact of the Lorentz
  collisions starts to become important for $\hat{\nu}_{ei} \gtrsim
  0.1(\sim|\omega|)$, and the two models start to diverge.  The use of
  the Lorentz operator leads to an overestimation of the impurity
  peaking factor in the TE mode case, when the collision frequency is high, 
  both in the poloidally symmetric case as well as in the asymmetric case.
  On the contrary, in the ITG mode case the use of the Lorentz operator leads 
  to an underestimation of the peaking factor. This is expected because of the 
  impact of $1 / \left(1 + i \nu_D\left(x\right) / \omega\right)$ in the 
  parallel dynamics term, which decreased in magnitude with increasing 
  $\nu_D\left(x\right)$. Since parallel dynamics decreases (increases) the 
  peaking factor for TE (ITG) modes we find an increase (a decrease) in the 
  peaking factor with increasing $\nu_D\left(x\right)$.
 }

\section{Conclusions}
\label{sec:conclusions}
The paper presents a quasilinear study of two collisionless TE mode
cases, driven by the density gradients, and the electron temperature
gradient, respectively, including their mode characteristics and their
effect on impurity transport.  Mode characteristics has been obtained
by linear gyrokinetic simulations using \gyro.  The poloidal wave
number was chosen as $k_\theta\rho_s=0.15$ for both cases to represent
the mode with the largest fluxes in nonlinear simulations.
In agreement with previous studies, the electron temperature gradient
driven mode is suppressed for small collisionalities, while the
density gradient driven mode \red{not only remains unstable even for
  very high collision frequencies, we also observe its transition to a
  dissipative TE mode}.
The dependence of growth rate \red{and real frequency} on safety
factor and magnetic shear is non-monotonic and within small
variations.

To investigate the peaking of high-$Z$ trace impurities in tokamak
plasmas we use an approximate \red{gyrokinetic}
model and compare it to results obtained with \gyro.  It is observed
that parameters such as $T_i/T_e$, $a/L_{Te}$ and $a/L_{ne}$ mainly
affect the peaking through their impact on mode characteristics,
particularly the factor $\omega_r/\left(\omega_r^2+\gamma^2\right)$ is
important to determine the effect of the impurity parallel
dynamics. \red{As noted before, in fluid modeling} this factor enters
directly into the parallel compressibility term of the approximate
model, and is consequently responsible for determining the size of
this contribution.  Parameters describing the magnetic geometry, $q$
and $s$, have a more significant influence on the peaking because of
their explicit appearance in certain contributions.  An increase in
magnetic shear typically leads to an increase of the impurity peaking
factor in the poloidally symmetric case, because of the increase in
the magnetic drift contribution.  However in the poloidally asymmetric
case \red{since the term describing the $\Ev\times \Bv$ drift of
  impurities in the non-fluctuating electrostatic potential has an
  explicit linear shear dependence, an increase in shear can lead to a
  significant reduction or enhancement of the impurity peaking,
  depending on the location of the potential minimum. }
Increasing safety factor leads to a decrease of the relative
significance of the impurity parallel dynamics contribution, but the
effect on the peaking depends on the sign of $\omega_r$ and for TE
modes, with $\omega_r > 0$, it results in an increase of the peaking
factor.  
\red{Nonlinear simulations
  in the density gradient driven TE case show only a very weak $q$
  scaling. This can be explained by a nonlinear shift in the poloidal
  wave number ($k_\theta \sim 1/q$, as shown in
  \cite{dannertjenko}). However, in the temperature gradient driven TE
  case, the $q$ scaling is even stronger non-linearly than the fixed
  $k_\theta$ linear modeling predicts. }
 
\red{ The model using the conservation properties of the full
  linearized collision operator for impurity self collisions show that
  collisions can only indirectly affect impurity transport through
  changes in the mode characteristics. We show that when a
  non-momentum-conserving model operator is used, such as a Lorentz
  operator, the parallel compressibility contribution to the peaking
  factor is modified leading to errors in the collisionality
  dependence. This effect becomes important when the impurity
  collision frequency becomes comparable to the mode frequency.}


\section*{Appendix A: Derivation peaking factor using a Lorentz collision operator}
\label{sec:appendixA}
\red{ In this section we derive a model for the impurity peaking
  factor similar to that represented by Eq.~(\ref{PeakingFactor}), but
  using the Lorentz collision operator instead of the linearized
  impurity-impurity collision operator.  We neglect $\ord(\epsilon)$
  corrections.  }

\red{ The Lorentz collision operator for impurity self-collisions is
  given by
\begin{equation}
C\left({g}_z\right)=\frac{\nu_D\left(x\right)}{2}\mathcal{L}\left({g}_z\right)
\equiv \frac{\nu_D\left(x\right)}{2}\frac{\p }{\p
  \xi}\left[(1-\xi^2)\frac{\p {g}_z}{\p \xi}\right],
\label{lorentzColl}
\end{equation}
where $\nu_D$ is the deflection frequency for self-collisions
$\nu_D\left(x\right)=\hat{\nu}_{zz}[{\rm Erf}(x)-G(x)]/x^3$, 
$\hat{\nu}_{zz}=n_z Z^4 e^4 \ln \Lambda/\left[4 \pi\epsilon_0^2
  m_z^{1/2}(2T_z)^{3/2}\right] $, and $\ln \Lambda$ is the Coulomb
logarithm.  $\rm Erf\left(x\right)$ is the error-function and
$G\left(x\right) = \left[{\rm Erf}\left(x\right) - x {\rm
    Erf}'\left(x\right)\right]/\left(2 x^2\right)$ the Chandrasekhar
function.  In the Lorentz operator $\xi=\xpa/x$ denotes the cosine of
the pitch-angle.  }

\red{
We assume the ordering
$\omega_{Dz} / \omega \sim \omega_{\ast z}^T / \omega \sim \omega_{E} / \omega \sim J_0(z_z) - 1 \sim 1/Z$,
and expand ${g}_z$ in $1/\sqrt{Z}$ keeping terms up to $\ord(Z^{-1})$,
i.e. ${g}_z \approx g_0 + g_1 + g_2$.  The $0^{\mathrm{th}}$ order
solution of  GK equation (Eq.~\ref{gke}) is
\begin{equation}
g_0 = \frac{Z e \phi f_{z0}}{T_z}
\label{eq:0thSol}
\end{equation}
with $C\left[{g}_0\right] = 0$. This, added to the adiabatic response,
$-Z e \phi f_{z0}/T_z$, merely tells that the impurities are so heavy,
and bound to the field lines through their high charge, that they do
not respond to electrostatic fluctuations to lowest order in $1/Z$. This
justifies neglecting the effect of impurities on the mode
characteristics, in spite of our assumption $n_zZ^2/n_e\sim 1$ (that
is required to make self-collisions dominate). }

The $1^{\mathrm{st}}$ order GK equation reads
\begin{equation}
  \frac{v_\parallel}{q R} \frac{\partial {g}_0}{\partial
    \theta}    
    -i\omega {g}_1 -
  C[g_1] = 0,
\label{eq:1st}
\end{equation}
and is solved by assuming that the solution can be written in terms of Legendre polynomials $P_n\left(\xi\right)$ as
${g}_1 = g_1^0\left(x\right) P_0\left(\xi\right) + g_1^1\left(x\right) P_1\left(\xi\right)$.
Here $P_0\left(\xi\right) = 1$, $P_1\left(\xi\right) = \xi$ and we remind about the properties
$\int_{-1}^1 d\xi P_n\left(\xi\right) P_m\left(\xi\right)=2\delta_{mn}/(2n+1)$
and
$\mathcal{L}\left[P_n\left(\xi\right)\right]=-n\left(n+1\right)P_n\left(\xi\right).$
The solution to Eq.~(\ref{eq:1st}) is found to be 
\begin{equation}
g_1 = - i \frac{v_\parallel}{q R}\frac{Z e f_{z0}}{T_z} \frac{\partial \phi}{\partial \theta} \frac{1}{\omega} \frac{1}{1 + i \nu_D / \omega}.
\label{eq:1stSol}
\end{equation}
Note that for a momentum conserving collision operator $C[g_1\propto
  \vpa f_{z0}]=0$ in Eq.~(\ref{eq:1st}), thus $\nu_D$ would not appear
in Eq.~(\ref{eq:1stSol}).

\red{  The $2^{\mathrm{nd}}$ order GK
  equation is
\begin{equation}
i \omega g_2 + C[g_2] = i \omega_{Dz} g_0 +  \frac{v_\parallel}{q R} \frac{\partial {g}_1}{\partial \theta} - i\frac{Z e f_{z0}}{T_z} \phi \left(\omega_{\ast z}^T + \omega \frac{z_z^2}{4}\right).
\label{eq:2nd}
\end{equation}
The velocity anisotropies enter in $\omega_{Dz}$ and $z_z$, which we can rewrite in terms of Legendre polynomials as 
$\omega_{Dz} \equiv (1/3) \left[ 2 P_0\left(\xi\right) + P_2\left(\xi\right) \right] \omega_{Dx}$
and
$z_z^2 \equiv (2/3) \left[P_0\left(\xi\right) - P_2\left(\xi\right)\right] z_x^2$,
where $P_2\left(\xi\right) = \left(3 \xi^2 - 1\right)/2$, and $\omega_{Dx} = \omega_{Dx}\left(x\right)$, $z_x = z_x\left(x\right)$ only depend on speed.
Furthermore, by noting that $\partial \energy/\partial \theta = 0$ (where $\energy = m_z v^2/2 + Z e \phi_E$) we can find the identity
\begin{equation}
\vpa \frac{\partial \vpa}{\partial \theta} = \frac{1}{3} \left(P_2\left(\xi\right) - P_0\left(\xi\right)\right) v^2 \frac{\partial \ln B}{\partial \theta} - \frac{Z e}{m_z} \frac{\partial \phi_E}{\partial \theta} P_0\left(\xi\right).
\label{eq:thetaDeriv}
\end{equation}
We now search for a solution to Eq.~(\ref{eq:2nd}) of the form 
\begin{equation}
{g}_2 = g_2^0\left(x\right) P_0\left(\xi\right) + g_2^2\left(x\right) P_2\left(\xi\right), 
\label{eq:g2expansion}
\end{equation}
where we realize that $g_2^2$ will not contribute to the particle flux since $\int_{-\infty}^{\infty}d\xi P_2\left(\xi\right)=0$. 
By substituting Eq.~(\ref{eq:g2expansion}) into Eq.~(\ref{eq:2nd}) and collecting the parts proportional to $P_0$ we find that
\begin{multline}
g_2^0 = \frac{Z e \phi}{T_z} f_{z0} \left[\frac{2}{3} \frac{\omega_{Dx}}{\omega} - \frac{z_x^2}{6} - \frac{\omega_{\ast z}^T}{\omega} + \frac{\omega_E}{\omega}\right] + \frac{1}{q^2 R^2} f_{z0} \frac{Z e}{T_z} \frac{\partial \phi}{\partial \theta} \frac{1}{\omega^2} \frac{1}{1 + i \nu_D / \omega} \frac{Z e}{m_z} \frac{\partial \phi_E}{\partial \theta}- \\- \frac{Z e }{T_z} \frac{f_{z0}}{q^2 R} \frac{v^2}{3 \omega^2} \frac{1}{1 + i \nu_D / \omega} \left[\frac{\partial}{\partial \theta} \left(\frac{1}{R} \frac{\partial \phi}{\partial \theta}\right) - \frac{1}{R} \frac{\partial \phi}{\partial \theta} \frac{\partial \ln B}{\partial \theta}\right].
\label{eq:2ndSol}
\end{multline}
From $g\approx g_0+g_1+g_2$, only the $g_2^0$ part of $g_2$ contributes to the particle
flux, thus the impurity peaking factor is found from solving
\begin{equation}
0 = \left\langle \Gamma_z \right\rangle = - \left\langle \frac{k_\theta}{B}\Im\left[\int d^3v J_0(z_z) g_z \phi^\ast\right] \right\rangle \approx - \left\langle \frac{k_\theta}{B}\Im\left\{\int d^3v \,g_2^0\, \phi^\ast\right\} \right\rangle,
\label{eq:PeakFactLorentz}
\end{equation}
where higher order than $1/Z$ corrections to the impurity flux are neglected.
}

\section*{Acknowledgments}
This work was funded by the European Communities under Association
Contract between EURATOM and {\em Vetenskapsr{\aa}det}. The
views and opinions expressed herein do not necessarily reflect those
of the European Commission. 
\newred{
The simulations were performed on resources 
provided by the Swedish National Infrastructure for Computing (SNIC) 
at PDC Center for High Performance Computing (PDC-HPC),
and on the HPC-FF cluster at the J\"ulich Supercomputing Center (JSC). 
}
The authors would like to thank J.~Candy
for providing the \gyro~code and Ye.~O.~Kazakov for fruitful
discussions.  \bibliographystyle{unsrt}

\end{document}